\documentclass[aps,pra,twocolumn,superscriptaddress]{revtex4-1}
\usepackage{CJK}

\usepackage{physics}
\usepackage{graphicx}
\usepackage{tikz}
\usepackage{subfigure}  
\usepackage{amsmath}
\usepackage{amsfonts}
\usepackage{amssymb}
\usepackage{bbm} 
\usepackage{pgfplots}

\usepackage{blindtext, xcolor}
\usepackage{comment}

\usepackage{color}

\begin{document}

\preprint{}

\begin{CJK*}{UTF8}{bsmi}
\title{Adiabatic preparation of squeezed states of oscillators and large spin systems coupled to a two-level system}

\author{Jiabao Chen (陳家葆)}
\author{Denis Konstantinov}
\affiliation{Quantum Dynamics Unit, Okinawa Institute of Science and Technology, Tancha 1919-1, Okinawa, 904-0495, Japan}

\author{Klaus M{\o}lmer}
\email[]{molmer@phys.au.dk}
\affiliation{Department of Physics and Astronomy, Aarhus University, Ny Munkegade 120, DK 8000, Aarhus C., Denmark}

\date{\today}

\begin{abstract}
We study a single two-level system coupled resonantly to an oscillator mode or a large spin. By adiabatically turning on a linear driving term on the oscillator or the spin, the eigenstates of the system change character and its ground state evolves into squeezed states of the oscillator or the spin. The robust generation of such states is of interest in many experimental systems with applications for sensing and quantum information processing.
\end{abstract}


\maketitle
\end{CJK*}
	
\section{Introduction}

The resonant transfer of excitation between a two-level system and a quantized bosonic mode occurs for a wide range of systems.
Among them are atoms in cavities (cavity QED)~\cite{Haroche1985,guerlin2010cavity}, cold ions in traps~\cite{blockley1992quantum, cirac1994quantum}, superconducting qubits (circuit QED)~\cite{wallraff2004strong}, electrons on liquid helium~\cite{dykman2003qubits, Lyon2006}, and many others.
Within the rotating wave approximation, the dynamics is described by the Jaynes-Cummings (JC) Hamiltonian~\cite{Jaynes1963}, and the discrete nature of the bosonic mode leads to interesting features
such as, the `collapse and revival' of Rabi oscillations, demonstrated by
atoms in microwave and optical cavities~\cite{Haroche1985, knight1982quantum, agarwal1984vacuum, thompson1992observation},
and the internal degree of freedom coupled with the center-of-mass motion of ions~\cite{blockley1992quantum,cirac1994quantum,agarwal1989collapse}. The non-linearity induced by the two-level system
causes effective Kerr non-linearities, which lead to squeezing~\cite{Carmichael1985c,meystre1982squeezed} and superpositions of coherent states of the oscillator~\cite{buvzek1992schrodinger}, and
similar to classical nonlinear systems, bistability and phase transitions are also present in the JC model
\cite{savage1988single,wang1996fokker}.

Squeezed states of an oscillator mode~\cite{plebanski1956wave, mollow1967quantum, Yuen1976, Henry1988} are non-classical states, whose fluctuations in one quadrature are smaller than in a coherent state. Squeezed states of light hold potential for high precision optical measurements, and squeezed microwave fields were recently shown to enhance the sensitivity in electron spin resonance experiments~\cite{bienfait2017magnetic}.
In this manuscript, we investigate the generation of squeezed states by adiabatic evolution of  a JC system subject to a slowly varying coherent drive on the oscillator component. This scheme is robust until it reaches  a critical driving strength. Here, the spectrum of the Hamiltonian collapses and for stronger driving the Hamiltonian displays a continuum of non-normalizable eigenstates.

We supplement the analysis of the oscillator with the study of a single two-level system coupled to a large spin, describing for example a central spin coupled to the collective symmetric states of other spin 1/2 particles in a spin-star configuration. Collective spin squeezed states
~\cite{wodkiewicz1985coherent, kitagawa1993squeezed} have non-classical correlations (entanglement) between their spin 1/2 constituents~\cite{janszky1990squeezing,sorensen2001entanglement}, and they have been proposed for use in precision clocks and magnetometers, and as entanglement resources for quantum information protocols~\cite{wasilewski2010quantum}. Spin squeezing of atomic ensembles may be obtained by suitably engineered interactions, and using Rydberg blockade interactions it has been proposed to use laser excitation pulses to implement adiabatic protocols to drive a large system of atoms into spin squeezed and entangled states,
~\cite{moller2008quantum,opatrny2012spin}. Here, we treat a case analogous to the JC model, namely of a classically driven spin interacting with a single two-level system, and we identify the states explored by this system under adiabatic variation of the interaction parameters. Unlike for the oscillator, the states of the large spin are always normalizable, but they evolve through spin squeezed and very non-classical quantum states.

The manuscript is structured as follows. In Sec. II, we introduce the Hamiltonian and we identify the analytical solution that is followed adiabatically by the system prepared in the ground state of the Jaynes-Cummings Hamiltonian, and subject to a gradually increased resonant driving of the oscillator mode until a maximum critical strength.
In Sec. III, we study the case of a two-level system coupled to an effective large collective spin, for which the initial ground state also transforms adiabatically through a sequence of eigenstates, and for which one may explore system eigenstates for all driving strengths. In Sec. IV we present our conclusions and an outlook.

\section{Jaynes-Cummings Hamiltonian with the oscillator subject to a resonant, linear drive}

\begin{figure}
	\includegraphics[]{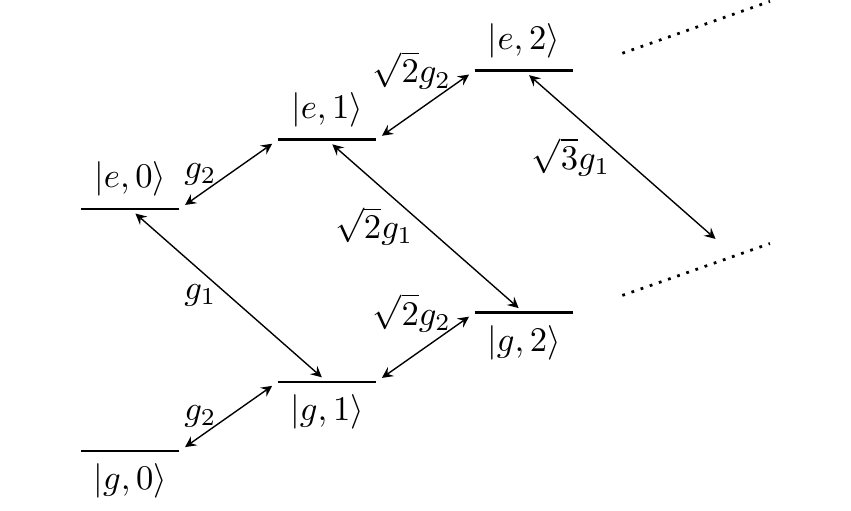}
	\caption{\label{fig-engergy-level}Level diagram showing the coupling between product states $|\sigma,n\rangle$, by JC type coupling with strength $g_1$ and $(a+a^\dagger)$ type coupling with strength $g_2$ }.
\end{figure}

We consider a two-level system (TLS) with a ground state $|g\rangle$ and an excited state $|e\rangle$, which is described by the Pauli raising and lowering operators, $\sigma_+ = \ketbra{e}{g}$ and $\sigma_- = \ketbra{g}{e}$.  It also interacts resonantly with strength $g_1$ with a quantized oscillator, described by the operators $a$ and $a^{\dagger}$, which in turn is subject to a resonant classical driving force of strength $g_2$. The schematic level diagram shown in Fig.~\ref{fig-engergy-level} depicts the product eigenstates $\ket{\alpha,n}=\ket{\alpha}\otimes\ket{n}$  of the uncoupled systems, where  $\alpha=g,e$ denotes the state of the TLS and $n$ is the excitation number of the oscillator mode.

In the rotating frame in the interaction picture (with respect to the bare atom and oscillator Hamiltonians) and with the rotating wave approximation, the Hamiltonian takes the form \cite{walls2007quantum} :
\begin{equation} \label{eq_1irwa}
	H_\text{I} =  g_1 \left(\sigma ^- a^{\dagger }+ \sigma ^+ a \right)+  g_2 \left(a^{\dagger }+a\right).
\end{equation}
A system of this kind can be implemented in a variety of quantum systems with two-level and oscillator degrees of freedom. In the case of a single trapped ion, $g_1$ can be obtained by driving a lower sideband optical transition, while $g_2$ can be implemented by an electric RF interaction with the charged particle motion. Atoms in cavities are excited by absorption of a cavity
photon with strength $g_1$, while resonant illumination of the cavity coherently excites the cavity mode with strength $g_2$ \cite{cirac1994quantum,blockley1992quantum}.

In the absence of pumping of the oscillator ($g_2 = 0$), the system is governed by the usual JC Hamiltonian and has the well known dressed eigenstates, $\frac{1}{\sqrt{2}}(|g,n\rangle \pm |e,n-1\rangle)$, with symmetric pairs of energies
$\pm g_1\sqrt{n}$ around the zero energy of the ground state $|g,0\rangle$. The presence of the $g_2$ term in the Hamiltonian \eqref{eq_1irwa} does not change the symmetry of the spectrum: for any eigenstate $|\Psi\rangle$  of $H_\text{I}$ with eigenvalue $E$, it is easy to verify that $(-1)^{a^\dagger a}|\psi\rangle$ is an eigenstate with eigenvalue $-E$.
All eigenvalues of the Hamiltonian (\ref{eq_1irwa}) for $g_2 < g_1/2$,
\begin{equation}\label{splitting}
		E_0 = 0, \ \ E_n^\pm = \pm \sqrt{n} g_1\left(1-\left(\frac{2g_2}{g_1}\right)^2\right)^\frac{3}{4}
\end{equation}
have been identified by Alsing and Carmichael together with expressions for the corresponding eigenstates \cite{Alsing1992c}.

\subsection{The eigenstates}

\begin{figure}
	\includegraphics[width=0.8\linewidth]{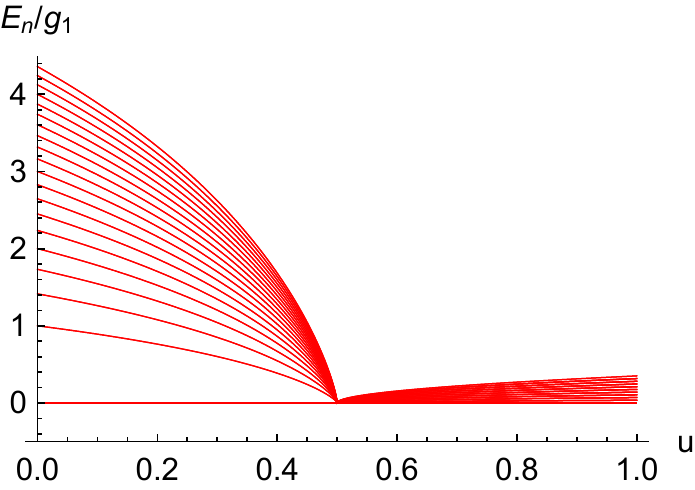}
	\caption{\label{pic_eigenvalues}
		A few smallest non-negative scaled eigenvalues: $E_n/g_1$, of the Hamiltonian \eqref{eq_1irwa}, in the truncated Fock basis of the oscillator ($N_\text{cut} = 2000$).
		The spectrum is symmetric with eigenvalues $\pm E$ around zero. The horizontal axis is a transition parameter
		$u\equiv 2g_2/(g_1+2g_2)$.
		}
\end{figure}

Diagonalizing the Hamiltonian over the full parameter range, we show the 21 lowest lying non-negative eigenvalues in  Fig.~\ref{pic_eigenvalues} as function of the parameter $u=2g_2/(g_1+2g_2)$.
The numerical results confirm the existence of the zero eigenvalue over the whole parameters range and the collapse of all eigenvalues to zero when $g_2 \rightarrow g_1/2$ ($u\rightarrow 1/2$).
In the limit where $u=1$, the system is only subject to the linear Hamiltonian $H_\text{I}= g_2(a+a^\dagger) = g_2\cdot x$, and the resulting, un-normalizable position eigenstates have a continuum of eigenvalues.
Due to the truncation in Fock space at $N_{cut}=2000$ in the numerical diagonalization, with a resulting largest $x_\text{rms} \simeq \sqrt{N_\text{cut}}$, however, the eigenvalues show almost equidistant spacing $\propto 1/\sqrt{N_\text{cut}}$ in the right hand side of the figure.

We now turn to the adiabatic evolution of the zero energy state as the driving strength $g_2$ is gradually increased. We expect to follow the zero energy eigenstate adiabatically as long as the change of Hamiltonian is slow compared to the energy gap to the other states, i.e., until we approach $g_2=g_1/2$. Although the full set of eigenstates is identified in \cite{Alsing1992c}, it is instructive to deal separately with the zero energy eigenstate, and we apply the product state ansatz $\ket{\Psi} = \ket{\chi} \otimes \ket{\phi}$ with the two-level systems parametrized as $\ket{\chi} = (\cos \theta/2 , \sin \theta/2)^T$. This yields an equation for each spin component of the eigenvalue equation, that must both be fulfilled by the oscillator state $\ket{\phi}$,
\begin{equation}\label{eq_linear_dependent}
\begin{split}
	\qty[g_1\sin \frac{\theta}{2} a^\dagger + g_2 \cos \frac{\theta}{2}  \qty(a + a^\dagger)] \ket{\phi}  = 0 \\
	\qty[g_1\cos \frac{\theta}{2} a + g_2 \sin \frac{\theta}{2}  \qty(a + a^\dagger)] \ket{\phi}  = 0.
\end{split}
\end{equation}
This system has unique solutions for $\ket{\phi}$ only if they are linearly dependent, which allows to find the relation
\begin{equation}
	\label{eq_sin_theta_2g}
	\sin \theta = -2g_2/g_1,
\end{equation}
with real solutions for $-\pi/2 < \theta < \pi/2$ as long as $\abs{g_2/g_1}<1/2$.
We note that the equations \eqref{eq_linear_dependent} for the oscillator state can be written in the familiar form
\begin{equation}\label{eq_bog}
	\qty(\mu a +\nu a^\dagger) \ket{\phi} = 0,
\end{equation}
where the real parameters
\begin{equation}\label{eq_mu_and_nu}
\begin{array}{l}
\nu = -\sin^2 (\theta/2) \sqrt{\sec(\theta)}\\
\mu = \cos^2 (\theta/2) \sqrt{\sec(\theta)}
\end{array}
\end{equation}
obey the normalization condition $\mu^2-\nu^2=1$.
The solutions to eq.~(\ref{eq_bog}) yield minimum uncertainty squeezed states \cite{Yuen1976, Henry1988} with variances in the $x=a+a^{\dagger}$ and $p=i(a^\dagger-a)$ oscillator quadratures given by
\begin{equation}
	\label{eq_variance}
	\begin{split}
		\mathrm{Var}(x) &= \expval{(a+a^\dagger)^2}= (\mu-\nu)^2= \sec\theta,\\
		\mathrm{Var}(p) &= \expval{(i(a^\dagger-a))^2}=(\mu+\nu)^2= \cos\theta.
	\end{split}
\end{equation}

By applying a classical field or a coherent drive to a cavity or a mechanical oscillator, coupled to a two-level system, the oscillator is driven adiabatically into a squeezed state, and we believe that this may be a robust, practical protocol to achieve appreciable squeezing. It is remarkable that while the adiabatically varying Hamiltonian passes between the Jaynes-Cummings Hamiltonian and the $x$ quadrature operator, and one might hence have expected the zero energy eigenstate to gradually transform into the $x=0$ position eigenstate, we instead observe strong squeezing of the conjugate observable $p$, as the system approaches the critical driving strength, $g_2=g_1/2$.

\subsection{The time evolved quantum state}

We have solved the time dependent Schr\"odinger equation under slow variation of the coupling strength $g_2$ and as one may expect, we find that beyond a finite degree of squeezing the system cannot follow the $E=0$ eigenstate adiabatically and the factorization in separate TLS and oscillator components fails. The fact that the Hamiltonian does not even have normalizable eigenstates as we explore values of $g_2 > g_1/2$ does not, however, prevent numerical solution of the  Schr\"odinger equation, and we have explored the dynamics within a truncated basis of harmonic oscillator states.

\begin{figure}
	\includegraphics[width=1.0\linewidth]{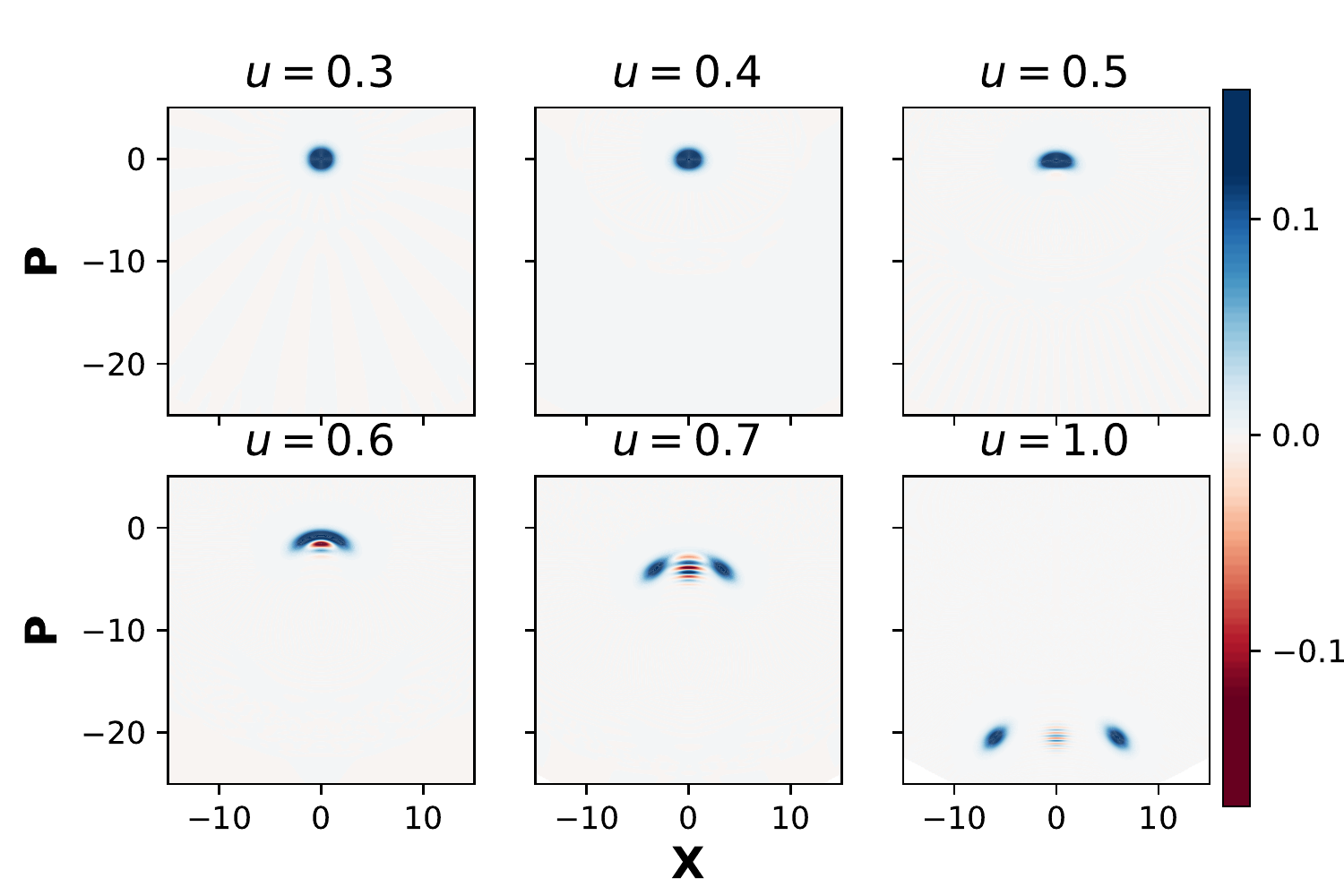}
	\caption[Time evolution]{\label{fig_time_evo}The time evolved system is initially able to follow the momentum squeezed eigenstates of the Hamiltonian adiabatically, but later non-adiabatic effects distort the phase space distribution and a 'Schr\"odinger cat-like' state appears when the Hamiltonian no longer supports discrete eigenstates. The simulation and Wigner function plot was generated by QuTip\cite{johansson2012qutip}.}.
\end{figure}

We assume the timedependent interaction strengths, 
\begin{equation}\label{eq_time_evo}
	g_1(t) =  (1-\frac{t}{T}) g,\ \ g_2(t) =  \frac{1}{2} \frac{t}{T}g,
\end{equation}
such that $u=2g_2/(g_1+2g_2)$ changes from $0$ to $1$ linearly in time. The results are obtained for a time scale $T=100 g^{-1}$ and a truncation of the oscillator Fock space at $N_\mathrm{cut}=2000$.
For early times the elliptic shaped Wigner function shows the graduate squeezing of the vertical $p$ component, cf., the upper panels of Fig.~\ref{fig_time_evo}
As we surpass $g_2=g_1/2$ the elliptic shape is distorted, and the non-adiabatic evolution gives rise to a `cat-like' superposition of components with well defined amplitude which are both displaced towards negative $p$ values and with negative quasi-probability 'fringes' along the $x=0$ line, cf. the lower panels of Fig.~\ref{fig_time_evo}.

\section{Two-level system coupled to an integer collective spin subject to a resonant, linear drive}

Let us consider the case of a two-level particle coupled to a collective spin $J$
\begin{equation} \label{eq_H_spin}
H_I=g_1\qty(\sigma^- J_+ + \sigma^+ J_-) + g_2\qty(J_+ + J_-),
\end{equation}
where we define $J_-\equiv J_x-i J_y$, $J_+ \equiv J_x + i J_y$.

Such a system may be implemented by the electron and nuclear spin in alkali atoms, and the large spin may also represent symmetric, collective states of a collection of two-level systems or the Schwinger representation of a pair of oscillators.
For any eigenstate $\ket{\Psi}$ of $H_I$ with eigenvalue $E$, $e^{-i\pi J_z}\ket{\Psi}$ is an eigenstate of $H_I$ with eigenvalue $-E$, so the spectrum is symmetric around zero as in the oscillator case.
For large J, the states close to the extremal $J_z$ eigenstate $|J,M=-J\rangle$, indeed, constitute an oscillator-like ladder and the weakly driven system shows similarities with the driven JC model.
When driven more strongly, we expect to see deviations from the JC dynamics, and due to the finite Hilbert space we obtain discrete, normalizable eigenstates of the Hamiltonian (\ref{eq_H_spin}) for all coupling parameters.

\begin{figure}
	\centering
	\includegraphics[width=\linewidth]{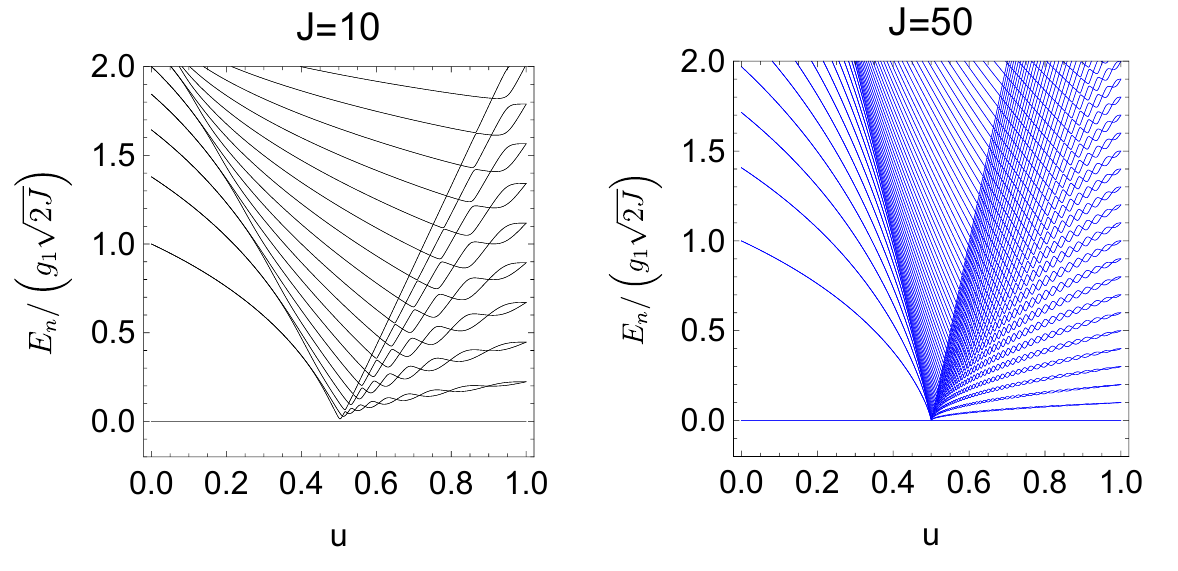}
	\caption{Lowest, non-negative energy eigenvalues for the two-level system coupled to a collective spin with $J=10$ (left) and $J=50$ (right). The scaling factor $(g_1\sqrt{2J})^{-1}$ ensures the similarity with Fig.~(\ref{pic_eigenvalues}) for the lowest eigenvalues for values of $u < 0.5$, where $u\equiv 2g_2/(g_1+2g_2)$.
	}\label{fig_j50}
\end{figure}

These properties are confirmed by numerical diagonalization of the Hamiltonian as illustrated for $J=10$ and $J=50$ in Fig.~\ref{fig_j50}.  When the eigenvalues are scaled by $\sqrt{2J}$, the lowest lying states for $g_2 < g_1/2$ show similar behavior as in Fig.~\ref{pic_eigenvalues}, while, for $g_2 > g_1/2$, the density of eigenstates depends on the finite number of angular momentum  states $2J+1$ rather than the numerical truncation of the oscillator system.

Systems with integer spin $J$ described by Eq. \eqref{eq_H_spin} have an odd number of distinctive eigenvalues, and due to the symmetry between positive and negative eigenvalues, there always exists eigenstates with eigenvalue 0. When we drive the system parameters across the transitional point we explore the transition between the zero energy eigenstates in the parameter ranges $g_2<g_1/2$ and $g_2>g_1/2$.

\subsection{The eigenstates}

 We will study the adiabatic evolution of the system when $g_2$ is gradually turned on, starting from the zero energy eigenstate $|g\rangle\otimes|J,-J\rangle$. The state will initially show features similar to the ones obtained in the previous section, but here we shall be able to understand the full dynamics through all parameter values from the behavior of the adiabatic eigenstates.

The similar structure of Eqs.\eqref{eq_1irwa} and \eqref{eq_H_spin} with the operator $a$ replaced by $J_-$ permit use of the same factorization Ansatz $\ket{\Psi}=\ket{\chi}\otimes\ket{\phi}$ to obtain the zero energy eigenstates, satisfying $H_I\ket{\Psi}=0$.

In the parameter range $0<g_2/g_1<1/2$, we can use the same expression for the TLS:
\begin{equation} \label{eq:chi1}
\ket{\chi} = (\cos(\theta/2),\sin(\theta/2))^\mathrm{T},
\end{equation}
where $\theta=\arcsin(-2g_2/g_1)$, such that \eqref{eq_H_spin} yields a single equation for $\ket{\phi}$,
\begin{align} \label{eq_spinbog2}
(\mu J_{-} + \nu J_{+}) \ket{\phi} &= 0
\end{align}
with $\nu/\mu = -\tan^2(\theta/2)$.
The solution of Eq.(\ref{eq_spinbog2}) is a minimum uncertainty spin squeezed state, also known as a generalized intelligent state~\cite{Rashid1978,Rashid1978a}, and it is known to be of the explicit form,
\begin{align} \label{eq:phi1}
	\ket{\phi}= C(\tau) e^{-\tau J_z} e^{i\frac{\pi}{2}J_x} \ket{J,0}_z=C(\tau) e^{-\tau J_z} \ket{J,0}_y
\end{align}
where $\tau=\log{\sqrt{\abs{\frac{\mu}{\nu}}}}=\log(\abs{\cot{\frac{\theta}{2}}})$ and $C(\tau)$ is a normalization factor to ensure $\braket{\phi}{\phi}=1$. We use $\ket{J,0}_{z(y)}$ to denote $J_{z(y)}=0$  eigenstates of the large spin.

The spin component uncertainties in the state $\ket{\phi}$ are

\begin{equation}\label{eq_uncertainty_spin_squeeze}
\begin{split}
\expval{\Delta J_y}& 	
=	\sqrt{\cos\theta} \sqrt{|\expval{J_z}|}/2\\
\expval{\Delta J_x}&	
= \sqrt{\sec\theta} \sqrt{|\expval{J_z}|}/2.
\end{split}
\end{equation}
where $\ev{J_z}=C^2_\tau \bra{J,0}_yJ_z  e^{-2\tau J_z}\ket{J,0}_y$ in our case.

For $g_2/g_1>1/2$, we may apply the complex argument solutions to $\theta = \arcsin(-2g_2/g_1)$, but for clarity we shall introduce an alternative parametrization with real arguments,
\begin{align} \label{eq:chi2}
		\ket{\chi}=\frac{1}{\sqrt{2}} \begin{pmatrix}e^{i\varphi/2}&, -e^{-i\varphi/2}\end{pmatrix}^\mathrm{T},\
\end{align}
where $\tan^2\varphi = 4 g_2^2/g_1^2 - 1,\ 0<\varphi<\pi/2$, resulting in the large spin equation
\begin{align} \label{eq_spinbog3}
	(e^{i(-\varphi-\pi/2)} J_{-} + e^{i(\varphi+\pi/2)} J_{+}) \ket{\phi} &= 0.
\end{align}
The pre-factors on $J_-$ and $J_+$ have the same absolute value, and for $g_2=g_1/2$, $\varphi=0$, and $\ket{\phi}$  is the $J_y=0$ eigenstate, while for larger $g_2$ and a finite $\varphi$, Eq.(15) describes an infinitely spin squeezed state with $M=0$ about an axis in the direction $\varphi$ with respect to the y-axis in the equatorial plane. As $g_2/g_1 \rightarrow \infty$, $\varphi$ approaches $\pi/2$ and $\ket{\phi}$ rotates towards the (expected) $J_x=0$ eigenstate of the large spin.
We note that these states have the explicit expression
\begin{align} \label{eq:infs}
	\ket{\phi}=  e^{i\varphi J_z}e^{i\frac{\pi}{2}J_x}\ket{J,0}_z = e^{i\varphi J_z}\ket{J,0}_y.
\end{align}
and that they may be attractive for precision measurements ~\cite{lucke2014detecting,sorensen2001entanglement}.

\subsection{Degeneracy of the $E=0$ eigenstates}

So far, we have disregarded an important fact in the description of the system: the energy eigenvalues for the Hamiltonian are two-fold degenerate for all values of the coupling strengths. This has the consequence that any weak coupling is sufficient to drive rotations of the state in the two-dimensional $E=0$ subspace and must be taken into account to properly describe the time evolution of the system, even if $g_1$ and $g_2$ change infinitely slowly. 

The degeneracy of the Hamiltonian eigenstates follows from the fact that
the Hamiltonian \eqref{eq_H_spin} commutes with the operator $R_x=\exp(-i\pi(J_x + \sigma_x/2))$, which applies a 180 degree rotation around the $x$ axis to both the two-level spin vector and the large angular momentum. This implies that for any eigenstate $\ket{\Psi}$ of $H_I$, $R_x\ket{\Psi}$ is also an eigenstate of $H_I$ with the same energy.

For integer values of $J$, $R_x^2 = e^{i(2\pi(1/2+J))}=-\mathbbm{1}$, and $i (-1)^J R_x$ thus has the eigenvalues $\pm 1$. It follows that assuming the zero energy product states $\ket{\Psi}=\ket{\chi}\otimes \ket{\phi}$ defined in the previous subsection, we can construct an orthonormal pair of joint eigenstates for $H_I$ and $i (-1)^J R_x$ (eigenvalues $\mp 1$),
\begin{equation}\label{eq_psipm}
\ket{\Psi_\pm}=
\frac{\ket{\Psi} \pm i (-1)^J R_x\ket{\Psi}}
{\sqrt{2} \sqrt{1 \pm \gamma}},
\end{equation}
where $\gamma$ is the real part of the state vector overlap $\gamma = \Re(\bra{\Psi} (iR_x(-1)^J\ket{\Psi}))$.
We shall denote these eigenstates $\ket*{\Psi_\pm^{(1)}}$ and $\ket*{\Psi_\pm^{(2)}}$ in the domains $0<g_2/g_1<1/2$ and $g_2/g_1>1/2$, respectively.

As the initial state $\ket{\Psi}$ for $g_2=0$ is orthogonal to $R_x \ket{\Psi}$ and hence $\gamma=0$, we can expand it as
\begin{equation}
\ket{\Psi} = \frac{1}{\sqrt 2}(|\Psi_+^{(1)}\rangle+|\Psi_-^{(1)}\rangle),\ \ \hbox{for $g_2=0$}.
\end{equation}

A general theory for adiabatic evolution with degenerate subspaces was presented in \cite{Wilczek1984}, but in our problem symmetry arguments suffice to obtain the approximate time dependent states (ignoring transitions to adiabatic eigenstates with non-vanishing energy). Linearity of quantum mechanics, and the fact that the time dependent $H_I$ commutes with $R_x$ and hence does not couple the $\ket{\Psi_\pm}$ eigenstates, ensures that the system will adiabatically evolve as the equal weight superposition of the symmetrized energy eigenstates, $\ket{\Psi} = \frac{1}{\sqrt 2}(|\Psi_+^{(1)}\rangle+|\Psi_-^{(1)}\rangle)$, as $g_2$ is slowly increased towards the value $g_2 = g_1/2$.

At $g_2=g_1/2$ a basis transformation to the eigenstates $|\Psi_\pm^{(2)}\rangle$  takes place. With our convention (\ref{eq:chi1},\ref{eq:phi1}) and (\ref{eq:chi2},\ref{eq:infs}) for the eigenstate $\ket{\Psi}$, we find that when approaching $g_2 = g_1/2$ from opposite sides, the limiting eigenstates obey the identities
\begin{eqnarray}\label{eq_chage_of_basis}
\ket*{\Psi_+^{(1)}} =\ket*{\Psi_+^{(2)}}\nonumber \\
\ket*{\Psi_-^{(1)}} =-i \ket*{\Psi_-^{(2)}},
\end{eqnarray}
where the first expression follows easily, while the second one requires a more careful analysis of the first order dependence of the states on the angle variable on either side of $g_2=g_1/2$.

Assuming that transitions to states with different energies are suppressed, we obtain the adiabatic approximation to the time dependent solution of the problem:
 \begin{equation}
 \ket{\Psi} =
 \left\{
   \begin{array}{ll}
     \frac{1}{\sqrt 2}\left( \ket*{\Psi_+^{(1)}}+\ket*{\Psi_-^{(1)}}\right), & \hbox{$g_2 < g_1/2$;} \\
     \frac{1}{\sqrt 2}\left( \ket*{\Psi_+^{(2)}}-i\ket*{\Psi_-^{(2)}}\right), & \hbox{$g_2 > g_1/2$.}
   \end{array}
 \right.
\end{equation}

\subsection{The time evolved quantum state}

\begin{figure}
	\includegraphics[]{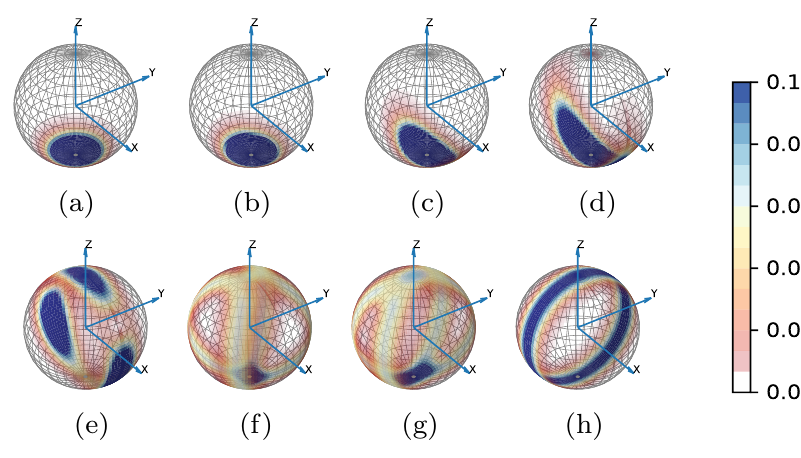}
	\caption{\label{fig_phi_sphere}
		The result of time evolution of the TLS and a large spin with J=20, subject to the Hamiltonian \eqref{eq_H_spin}, with time varying coefficients $g_1(t)$ and $g_2(t)$ as in the previous section. The result is represented by the Husimi Q-function of the reduced density matrix of the large spin $\rho_\phi$, shown for different values of $u\equiv 2g_2/(g_1+2g_2)$.
		While increasing $g_2$ and decreasing $g_1$, the collective spin subsystem starting form $\ket{J,-J}$ ($u=0$), evolves along spin squeezed states and spin eigenstates determined by the analytical arguments and expressions in the main text. (a) u=0.1; (b) u=0.4; (c) u=0.49; (d) u=0.499; (e) u=0.501; (f) u=0.6; (g) u=0.7 (h) u=1.0.
	}
\end{figure}

We have numerically tested the validity of the restriction of the dynamics to the $E=$ subspace and the formation of superposition states in this subspace. Using the temporal ansatz (8),  with a duration $T > 10000 g^{^-1}$ we find good agreement throughout the entire time evolution between the numerical solution of the time dependent Schr\"odinger equation and our analytical eigenstate expressions.
Fig. 5 shows results obtained with $T= 50000 g^{-1}$, and we observe how the state occupied for $g_2 < g_1/2$ first develops into a highly spin squeezed state in the $xz$-plane and then a $J_y=0$ eigenstate as shown by the Husimi Q-function in the upper panels in Fig.~\ref{fig_phi_sphere}. For $g_2 > g_1/2$ the numerical solution reveals an intricate patterns of two vertical rings that rotate in opposite direction, cf., the lower panels in Fig.~\ref{fig_phi_sphere}. These rings are the $M=0$ eigenstate components (16) with opposite angular argument $\varphi$, populated simultaneously in Eq.(20) and finally coalescing into the $J_x=0$ eigenstate.  The simultaneous occupation of two differently oriented spin squeezed states is similar to the observation in Fig.3 of the progression from a momentum squeezed state of the time evolved harmonic oscillator into two position squeezed wave packet components.
In the limit of $g_2 \gg g_1$, when the two angular momentum states coincide in the $J_x=0$ eigenstate, the TLS occupies a superposition of $\ket{\chi_1}$  and $\ket{\chi_2}$, forming the TLS ground state $\ket{g}$.

\begin{figure}
	\includegraphics[width=0.99\linewidth]{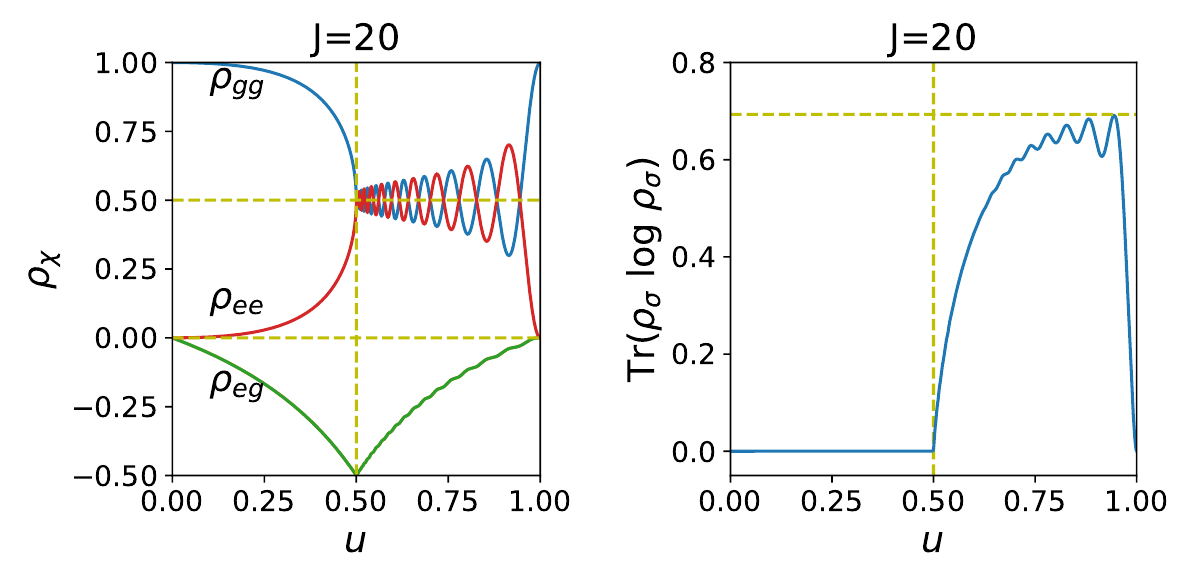}
	\caption{The result of slow time evolution by the Hamiltonian \eqref{eq_H_spin} illustrated by the reduced density matrix elements of the two-level subsystem $\rho_\chi(t)$ and its von Neumann entropy. The horizontal axis is the transition parameter $u\equiv 2g_2/(g_1+2g_2)$.}
	\label{fig_rhoaelements}
\end{figure}

To further illustrate how the system evolves from a product state to an entangled superposition state and back to a product state, we shown the time evolution of the reduced density matrix of the TLS subsystem, $\rho_\chi(t)$,  in the left panel of Fig.~\ref{fig_rhoaelements},  corresponding to the Husimi Q-Function shown in Fig.~\ref{fig_phi_sphere}.
The rapid oscillation between density matrix element $\rho_{gg}$ and $\rho_{ee}$, is caused by interference terms in the scalar product $\langle \phi|e^{-i\pi J_x}|\phi\rangle$ of the two $M=0$ state with respect to the rotated axes and is only reproduced correctly by  the analytical $E=0$ superposition states  Eq.(20) if the proces duration is longer than $10000 g^{^-1}$.
The right panel in Fig.~\ref{fig_rhoaelements} shows the von Neumann entropy of the state of the TLS, confirming the emergence and disappearance of entanglement of the joint quantum state of the system.

\section{Discussion}


To summarize, we have analyzed the dynamics of a  two-level system coupled resonantly to an oscillator and to a large spin. We have shown that factorized zero energy states exist under the resonant driving of the oscillator or large spin, and that the adiabatically evolved state becomes squeezed and entangled as the driving amplitude is gradually increased. These results supplement related ideas for generation of squeezed and non-classical states in the literature ~\cite{moller2008quantum, opatrny2012spin} and due to the generic Hamiltonians assumed in this work  they may inspire experimental protocols for squeezing of field and motional oscillators and collective spins in a variety of quantum systems.

Our method of solution may go well beyond the Hamiltonians studied in this article and apply to the coupling of a TLS and any ancillary system with a Hermitian adjoint pair of operators $K^\dagger$ and $K$,
\begin{equation}\label{eq_def_Hg}
\begin{split}
H_I \equiv
g_1 \left(\sigma ^- K^{\dagger }+ \sigma ^+ K \right)+  g_2 \left(K^{\dagger }+K\right).
\end{split}
\end{equation}
The similarity with Eqs.~(\ref{eq_1irwa},\ref{eq_H_spin}) invites use of the product state ansatz: $\ket{\Psi} = \ket{\chi)} \otimes \ket{\phi}$ for an $E=0$ eigenstate, where $\ket{\chi} = (\cos \theta/2 , \sin \theta/2)^T$, leads to two equations
\begin{equation}\label{eq_linear_dependent2}
\begin{split}
\qty[\sin \frac{\theta}{2} K^\dagger + \frac{g_2}{g_1} \cos \frac{\theta}{2}  \qty(K + K^\dagger)] \ket{\phi}  = 0 \\
\qty[\cos \frac{\theta}{2} K + \frac{g_2}{g_1} \sin \frac{\theta}{2}  \qty(K + K^\dagger)] \ket{\phi}  = 0,
\end{split}
\end{equation}
and with the constraint $\sin\theta = -2 g_2/g_1$, we find that $\ket{\phi}$ must solve the equation,
\begin{equation}\label{eq_bog3}
\qty(\mu K +\nu K^\dagger) \ket{\phi} = 0,
\end{equation}
where $\nu/\mu=-\tan^2(\theta/2)$.

While lower sideband excitation of a trapped ion can be used to implement the JC model, upper sideband excitation leads to the so-called  anti-JC model with interaction $\propto (\sigma^- a + \sigma^+ a^\dagger)$~\cite{mekkhof1996generation}, and simultaneous driving of both sidebands implement ladder operators among squeezed Fock states (Bogoliubov transformed linear combinations of $a$ and $a^\dagger$) \cite{kienzler2017quantum} with Hamiltonians that are also explicitly of the form of \eqref{eq_def_Hg}. To obtain similar results as in the present article, we recall that one must verify the existence of the $E=0$ product state which will depend on the properties of the ancillary system (e.g., it does not occur for Eq.(9) with a half integer spin) and of the operator $K$.

Further applications may go well beyond quantum optics as, e.g., Guti\'{e}rrez-J\'{a}auregui and Carmichael  \cite{gutierrez2018quasienergy} have emphasized the interesting formal equivalence between the driven Jaynes-Cummings Hamiltonian and the Dirac Hamiltonian of a charged particle subject to an external electromagnetic field and where a similar transition between discrete and continuous spectra appears.

\begin{acknowledgments}
 We thank Prof. Thomas Busch for helpful discussions and advise on the manuscript.
 The authors acknowledge support from Villum Foundation and Okinawa Institute of Science and Technology Graduate University.
\end{acknowledgments}

\bibliography{extracted}

\end{document}